\documentclass[aip,numerical]{revtex4-1}

\usepackage{graphicx}
\usepackage{dcolumn}
\usepackage{bm}
\usepackage{amsfonts}
\usepackage{amsmath}
\usepackage{amssymb}
\usepackage{amstext}
\usepackage{latexsym}
\usepackage{textcomp}
\usepackage{color}
\usepackage{tikz}
\usepackage{natbib}



\begin{document}
\def\myfrac#1#2{\frac{\displaystyle #1}{\displaystyle #2}}

\title{Femtosecond laser crystallization of amorphous titanium oxide thin films}
\author{Jan S. Hoppius}
\email[Corresponding author: ]{hoppius@lat.rub.de}
\affiliation{Ruhr-Universit\"at Bochum, Chair of Applied Laser Technology,
Universit\"atsstra\ss e~150, 44801 Bochum, Germany}
\author{Danny Bialuschewski}
\email[Corresponding author: ]{danny.bialuschewski@uni-koeln.de}
\affiliation{Institute of Inorganic Chemistry, University of Cologne, Greinstrasse 6, 50939 Cologne, Germany}
\author{Sanjay Mathur}
\email[Corresponding author: ]{sanjay.mathur@uni-koeln.de}
\affiliation{Institute of Inorganic Chemistry, University of Cologne, Greinstrasse 6, 50939 Cologne, Germany}
\author{Andreas Ostendorf}
\email[Corresponding author: ]{andreas.ostendorf@ruhr-uni-bochum.de}
\affiliation{Ruhr-Universit\"at Bochum, Chair of Applied Laser Technology,
Universit\"atsstra\ss e~150, 44801 Bochum, Germany}

\author{Evgeny L. Gurevich}
\email[Corresponding author: ]{gurevich@lat.rub.de}
\affiliation{Ruhr-Universit\"at Bochum, Chair of Applied Laser Technology,
Universit\"atsstra\ss e~150, 44801 Bochum, Germany}

\date{\today}

\begin{abstract}

In this paper we demonstrate experimentally that crystalline phases appear in amorphous titanium oxide upon processing with ultrafast laser pulses. Amorphous titanium thin films were produced by plasma-enhanced chemical vapor deposition (PECVD) and exposed to femtosecond laser pulses. Formation of rutile phase was confirmed by X-ray diffraction, Raman measurements and electron backscattering diffraction. A rang of processing parameters for the crystallization is reported and possible background mechanisms are discussed.

\end{abstract}



\maketitle


Ultrafast (femtosecond and picosecond) lasers are used for material processing e.g., for surface structuring and chemical conversion \cite{Bauerle} or for shock wave peening \cite{Jyotsna,JanSWP}. Micromachining of transparent materials with ultrashort laser pulses finds a growing number of applications in photonics, medicine and engineering \cite{Gattass,Feng}. These applications exploit the fact that the energy is coupled to the surface and triggers the following processes at a very short time scale. Thermal diffusion plays minor role and cannot deteriorate the high resolution of optical focusing by heat conducting around the processed spot. Hence ultrafast lasers are well established for cutting and structuring of thin films. However, they are supposed to be not suitable for processes like e.g., annealing, alloying or crystallization, in which the temperature has to evolve slowly to facilitate diffusion of heat and matter. Nevertheless surface alloying of metals with ultrafast lasers was recently demonstrated numerically \cite{Derek} and experimentally \cite{Melting}. The surprising aspect of femtosecond surface alloying is that on one hand, the alloying requires slow temperature evolution and long molten phase to facilitate diffusion of atoms. On the other hand, the metal surfaces remain molten upon single femtosecond pulse illumination during approximately $0.1 - 1$\,ns, which is not sufficient to form measurable alloyed layers. This contradiction is solved by the heat accumulation upon processing with multiple pulses: In order to facilitate diffusion, the molten phase must be extended by increasing either the pulse overlapping rate \cite{Melting} or the laser repetition rate \cite{Gamaly,Weber}. 

Crystalline structure modification by lasers is usually done in two different ways: (1) if a disordered crystalline structure is needed, the laser interaction on the sample must be as short as possible in order to achieve highest temperature gradients in time and space. Accordingly, the laser-induced amorphisation of silicon was reported for laser processing with picosecond \cite{Liu1979} and femtosecond \cite{Bonse2002} lasers, as well as a side effect of laser drilling with UV femtosecond laser pulses \cite{AmSi}. (2) if a highly-ordered crystalline structure should be achieved, crystallization has to happen slowly, to give atoms enough time to find optimal positions in the lattice. Long-pulsed \cite{Coucheron} or continuous wave \cite{SiCW,Sera} lasers are used to crystallize amorphous silicon in thin films or in optical fibers. However, waveguides can be written in transparent materials also with ultrafast lasers \cite{Davis:96,Miura1997,Nolte2003}. In this case the refractive index changes due to the laser-induced thermal shock \cite{Feng,Davis:96,Bharadwaj,Glezer1997} and following structural rearrangement \cite{Chan:01} or due to recrystallization \cite{Komatsu,Stone2015}. Ultrafast laser-induced crystallization is also reported for femtosecond laser ablation of highly-absorbing amorphous silicon layers (a-Si:H) for solar cells \cite{Choi,Nayak2007}. Laser annealing of semiconductors with ultrashort pulses can be explained by non-thermal electron-hole plasma, which weakens the lattice bounds and distorts the crystalline lattice \cite{Sundaram2002}.

In this paper we demonstrate that amorphous titanium oxides ($TiO_x$) can be converted to crystalline oxides ($TiO_2$) also upon irradiation with highly overlapped femtosecond laser pulses without any ablation and suggest an explanation of this effect. The results reported here develop thin film laser processing technology and methods for manufacturing thin-film photonic and electronic devices, which currently require at least two different lasers: one (femtosecond laser) for cutting and selective ablation, and one (continuous or long-pulse laser) for recrystallization of the amorphous films. We demonstrate that both these technological problems can be solved with only one femtosecond laser with high pulse repetition rate. Moreover, the femtosecond laser recrystallization is the only method providing highly localized heat treatment and hence is applicable for high-resolution processing of thin films. It enables recrystallization of films, which melting point is much higher than that of the substrate. Recent publications show that such treatment of titania and iron oxide photoanodes increases the efficiency of photo-electrochemical water splitting performance by more than two times \cite{AdvEngMat}.

\begin{figure}
\begin{center}
\includegraphics[width=8cm]{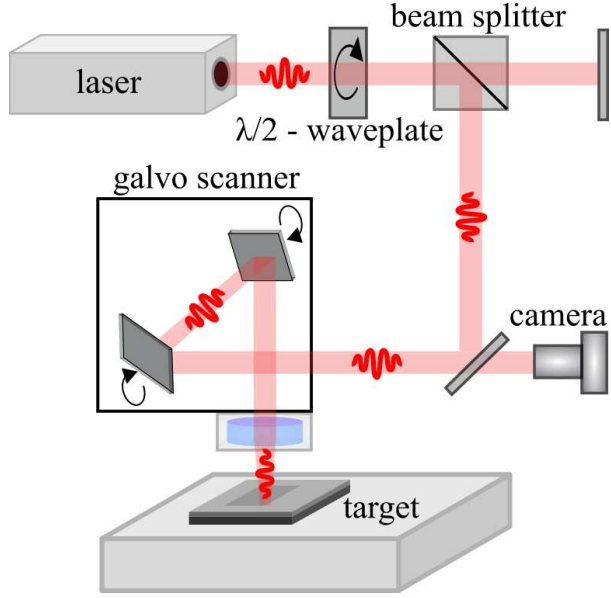}
\end{center}
\caption{Schematic representation of the experimental setup and the sample structure.}
\label{setup}
\end{figure}

The experimental setup is schematically shown in fig.~\ref{setup}. Two femtosecond lasers (either {\it Tangerine} produced by Amplitude Systems, pulse duration $\tau_p\approx280\,$fs, repetition rate $f=2000\,$kHz, central wavelength $\lambda=1030\,$nm; or {\it Spitfire Ace} produced by Spectra Physics, pulse duration $\tau_p\lesssim 100\,$fs, line width $\Delta\lambda=60\,$nm, repetition rate $f=5\,$kHz, central wavelength $\lambda=800\,$nm) were used in the experiments to cover a broader range of processing parameters. The linear-polarized laser beam first passes through a half-wavelength plate ($\lambda/2$-wave plate) and a polarizing beam splitter 
to adjust the pulse intensity. Afterwards, the beam is guided to the galvanometer scanner ({\textit{ScanCube 7 / 10}, produced by ScanLAB), which is used to raster the laser spot and to focus it onto the sample surface, the spot size was $d\gtrsim10\,$\textmu{}m. 

The titanium oxide layers were prepared by plasma-enhanced chemical vapor deposition (\textit{Type Domino} by Plasma Electronic GmbH) on aluminium substrates at an RF power of $100$ W with $20 $ sccm oxygen for 60 minutes using titanium-tetraisopropoxide [Ti(O$^i$Pr)$_4$] at 80\textdegree C precursor temperature.  Due to the nature of the PECVD process, only amorphous oxides are deposited. As the deposited titanium oxide has no defined structure, the amorphous phases are denoted as $TiO_x$.

The samples were processed in air at room temperature at different average laser powers to investigate the dependence of oxide modification process on the laser parameters such as fluence and repetition rate. The crystalline structure of the samples was studied by Raman microscopy ({\it Renishaw inVia} Raman microscope, excitation wavelength 633\,nm), XRD and electron backscattering diffraction.


\begin{figure}
\begin{center}
\includegraphics[width=8cm]{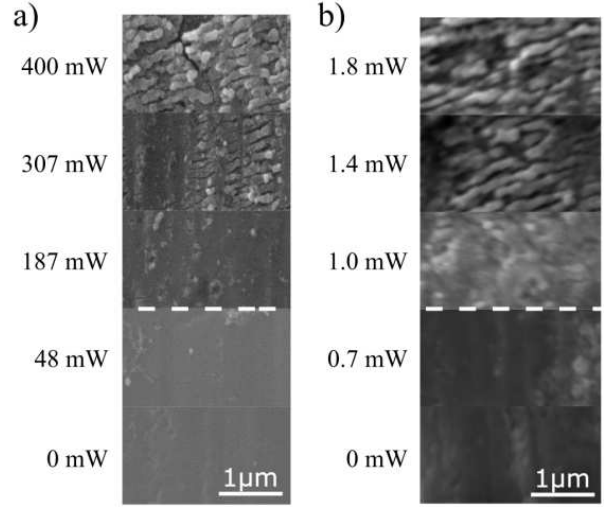}
\end{center}
\caption{Titanium oxide samples processed with femtosecond pulses of different pulse energies at a) 2000 kHz and b) 5 kHz repetition rate. At low pulse energies, minor surface defects occur, which change into quasi-periodic ripples with increasing energy. Images marked as "0 mW" correspond to not-processed samples.}
\label{SEM_TIOX}
\end{figure}

First we describe experiments with $TiO_x$ samples and high-overlapping rate laser. The deposited titanium oxide builds a homogeneous layer with a thickness of $1000\,nm$ on a $1\,mm$ thick aluminium substrate. The scanning speed in the experiments was $v=280$\,\textmu{}m/s, which means that each site on the sample surface was exposed to $N=df/v\approx7\times10^4$\,pulses in each line with time interval of $\delta t_1=f^{-1}=0.5\,$\textmu{}s between them. The laser-processed area was covered with 1000 lines per millimeter (the line length was $L=1\,$cm), i.e, after $\delta t_2=L/v\approx36\,$s the next line exposed the same site with the next portion of $N\approx7\times10^4$\,pulses. In total, each site is exposed to approximately ten such laser-heating cycles. Different areas were processed with increasing pulse energy from $E_{min}=60\,$nJ to $E_{max}=200\,$nJ per pulse (corresponding to the average laser power from $P=120\,mW$ to $P=400\,mW$).

The experiments with high-repetition rate {\it Tangerine} laser demonstrated conversion of amorphous $TiO_x$ to rutile. Starting from approximately $E_p=100\,$nJ per pulse, a slight change in the color of the sample surface was observed, which can indicate an increase in the refractive index of the layer.

\begin{figure}
\begin{center}
\includegraphics[width=16cm]{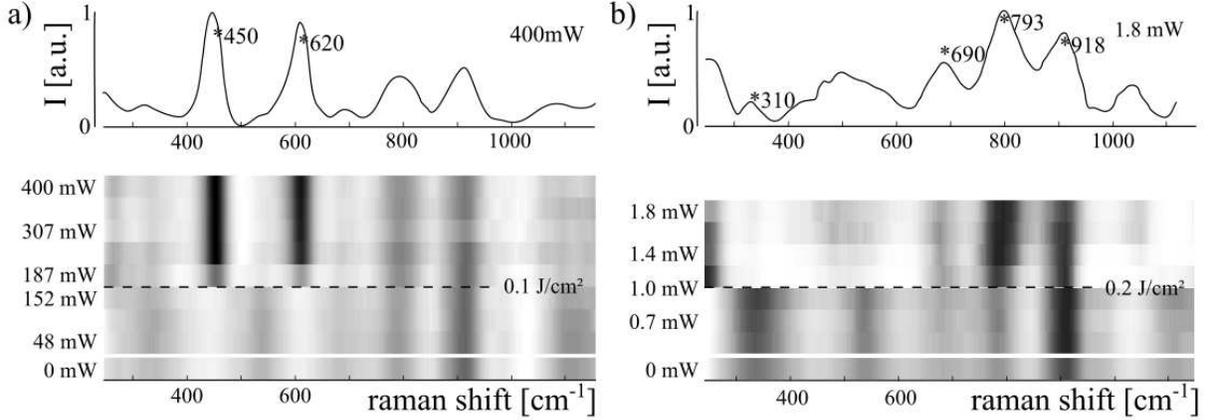}
\end{center}
\caption{Raman measurement of $TiO_x$ samples after femtosecond laser processing at a) 
$f=2000\,kHz$ and b) $f=5\,kHz$. The Raman signal changes at a threshold fluence of 0.1 $J/cm^2$ and 0.2$J/cm^2$. Only for high repetition rates characteristic rutile peaks at $\lambda^{-1}\approx 450\,cm^{-1}$ and $\lambda^{-1}\approx 610-620\,cm^{-1}$ are visible.}
\label{raman}
\end{figure}

Characteristic rutile peaks at $\lambda^{-1}\approx 450\,cm^{-1}$ and $\lambda^{-1}\approx 610-620\,cm^{-1}$ in the Raman spectrum were observed in the range of incident average powers from $P\gtrsim 200\,mW$ to the maximal power used in the experiments $P\approx 400\,mW$ (corresponding to pulse fluences from $F\approx0.1\,J/cm^2$ to $F\approx0.2\,J/cm^2$), see Figure~\ref{raman}. The broadening of the peaks and the red shift of the peak at $635\,cm^{-1}$ (which is detected at $620\,cm^{-1}$) can be explained by the small size of the nanocrystalline structure of the laser-crystallized films \cite{Barborini}.

The crystallization of amorphous $TiO_x$ samples was also confirmed by x-ray diffraction (XRD) measurements. In case of $TiO_x$ on aluminium (see Fig \ref{xrd}), the three most intense signals belong to the aluminium substrate, while the very broad neighboring peaks are different aluminium oxides, which formed by laser treating the substrate under ambient conditions. The peaks around 12.5, 16.4, 24.3, 25.2, 30.2\textdegree{} can be assigned to the (110), (101), (211), (220) and (301) planes of $TiO_2$ rutile phase, respectively. This phase usually forms around 600\textdegree C during thermal crystallization \cite{Hanaor}, thus XRD confirms the Raman measurements. Electron backscattering diffraction (EBSD) of laser-processed $TiO_x$ sample (see Fig.~\ref{rutile_ebsd}) confirms that the average size of the rutile nanocrystallites is approximately $0.7\pm 0.2\,\mu$m.

It is interesting to note that application of a low-repetition rate laser ({\it Spitfire}, $f=5\,kHz$) even in a broader range of pulse fluences (up to $P=1.8\,mW$, which corresponds to $F\approx0.4\,J/cm^2$) to the same sample was less successful. We observed changes in the Raman Spectrum (see figure ~\ref{raman}b), but these peaks do not correspond to any well-defined crystalline structure of titanium oxides. Thus, no crystallization with the low-repetition rate laser was detected by Raman and XRD measurements for $TiO_x$ layers deposited on aluminium substrate, even though aluminium is a good reflector for the laser wavelength and the laser energy absorbed in the film increased due to the double pass of the laser radiation through the target.

\begin{figure}
\begin{center}
\includegraphics[width=8cm]{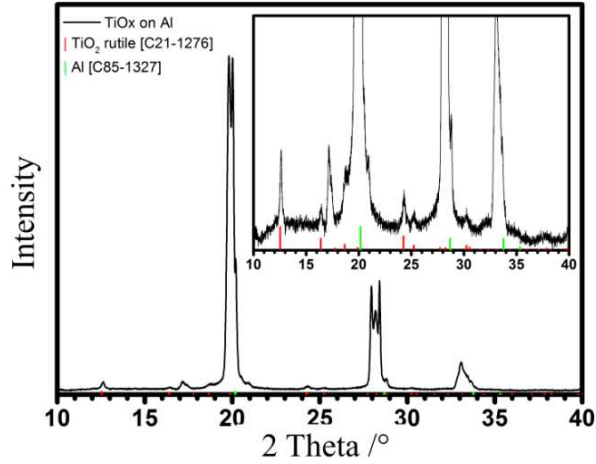}
\end{center}
\caption{XRD measurement of $TiO_x$ sample after laser processing. Characteristic peaks for rutile can be observed at the angles $2\theta\approx12.5,\, 16.4,\, 24.3,\, 25.2,\, 30.2$ degrees. The inset corresponds to the zoomed in spectrum to show small peaks. The positions of the reference peaks are marked with vertical lines.}
\label{xrd}
\end{figure}

\begin{figure}
\begin{center}
\includegraphics[width=8cm]{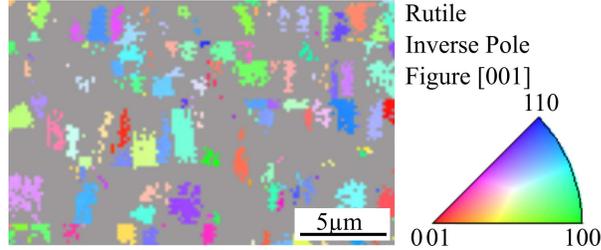}
\end{center}
\caption{Electron backscattering diffraction measurements were carried out with Zeiss LEO 1560VP SEM with DigiView 4 EBSD detector. The average crystal size is $0.7\pm 0.2\,\mu$m, the sample roughness was 300\,nm.}
\label{rutile_ebsd}
\end{figure}

The experimental findings can be explained under assumption that for crystallization, the surface temperature should not decrease to the initial value between two consequent incident laser pulses, so that pulse-to-pulse temperature accumulation happens. In this case the characteristic average cooling and heating times are controlled by the laser repetition rate and the velocity of the laser spot on the surface, as it is for cw lasers. If no material ablation takes place, the heat is removed from the laser-heated spot in two ways: into the bulk of the substrate and along the surface through the deposited oxide layer. The second mechanism can be neglected due to low oxide thickness, thus, the heated depth of the substrate during the time between two pulses $\delta t_1$ can be estimated as $L_{th}\sim\sqrt{Dt_1}$ with $D$ - the thermal diffusivity of the substrate. The temperature increase can be estimated by balancing the incident laser pulse energy and the energy needed to heat up the affected substrate volume. The latter is limited by the laser spot size and the heated depth $L_{th}$, if this one is smaller or comparable to the laser spot diameter $d$. In this case with the incident laser fluence (energy per area) $F$ and the specific volumetric heat capacity $c$ we obtain $\Delta T\approx F/cL_{th}$. If $L_{th}\gg d$, the affected substrate volume can be estimated as a half-sphere of the radius $L_{th}$, and the pulse-to-pulse temperature rise is by the factor $(d/L_{th})^2$ smaller: $\Delta T\approx Fd^2/cL^3_{th}$.

Now we can estimate the temperature rise for $TiO_x$ on aluminium ($D=10^{-4}\,m^2/s$, $c=2.4\,J/(cm^3K)$) substrate.
The wavelength and the applied peak power (and fluence) for the processing by the {\it Tangerine} laser are similar to that of the {\it Spitfire} laser, so we estimate for the absorbed fluence $F=0.1\,J/cm^2$. 
In the experiments with the high-repetition rate {\it Tangerine} laser the interval between the pulses ($\delta t_1=0.5$\,\textmu{}s) is enough for the heat to diffuse over the distance of $L_{th}\approx 7\,\mu m$ and the corresponding pulse-to-pulse temperature rise is $\Delta T\approx 60\,K$. 

In case of the low-repetition rate {\it Spitfire} laser, each site is exposed to only $N\approx 200$ pulses and the heated depth can be estimated as $L_{th}\approx 140\,\mu m$. This is  much larger than the spot diameter and the temperature rise is only $\Delta T< 0.1\,K$. Even if we suppose linear temperature accumulation, the total largest possible increase in the surface temperature $N\times\Delta T$, it will be far below the melting point of titanium dioxide $T_m=2100\,K$. However if quartz substrate is used ($D=10^{-6}\,m^2/s$ and $c=1.9\,J/(cm^3K)$), due to the low thermal diffusivity the heated depth ($L_{th}=16\,\mu m$) is comparable with the spot diameter and the temperature increase per pulse is $\Delta T\approx 30\,K$ for a pulse fluence of $F=0.1\,J/cm^2$.


In conclusion, we have demonstrated that the amorphous titanium oxide can be crystallized upon femtosecond laser radiation. The explanation of this effect can be given based on the following processes: Although after each pulse the laser-triggered surface temperature increases on the picosecond time scale, the limited thermal conductivity of the substrate confines the heat within a thin diffusive layer. As the next laser pulse arrives, the sample had not enough time to cool down to the initial temperature of the surrounding media and the heating cycle starts again, but from a higher temperature \cite{Gamaly}.
In this way there are two timescales of the temperature change: (1) the scale of tens of picoseconds, which is natural for single femtosecond pulse processing. (2) timescale which is relevant for crystallization, defined by the heat accumulation between the following laser pulses. This timescale can be controlled by the thermal diffusivity of the materials, pulse repetition rate, the number of pulses and the diameter of the laser spot. If the spot diameter is larger or comparable to the diffusion length, the heat diffusion is one-dimensional (only in the depth of the substrate) and the cooling is slower. For $L_{th}\gg d$, the heat diffuses effectively in all directions into the substrate and the cooling velocity increases. Effective crystallization requires low cooling rates, so $L_{th}\ll d$ is advantageous. 

\section*{ACKNOWLEDGMENTS}
The authors acknowledge the financial support by the DFG in the frame of SPP 1839 {\it Tailored disorder}, the project {\it EnLight} and Norbert Lindner (Ruhr-Universit\"at Bochum) for help with EBSD measurements.

%

\end{document}